\documentclass{sig-alternate-05-2015}

\usepackage[T1]{fontenc}
\usepackage{times}
\usepackage{todonotes}
\usepackage{mathtools}
\usepackage{enumitem}
\usepackage{listings}
\usepackage{hyperref}
\usepackage{mdframed}
\usepackage{caption}
\usepackage{array}
\usepackage{tabu}
\usepackage{booktabs}
\usepackage{algorithm}
\usepackage{algpseudocode}
\usepackage{microtype}

\usepackage{graphics}

\def\sharedaffiliation{
\end{tabular}
\begin{tabular}{c}}

\newcommand{\code}[1]{\mbox{\texttt{#1}}}

\algnewcommand\algorithmicforeach{\textbf{for each}}
\algdef{S}[FOR]{ForEach}[1]{\algorithmicforeach\ #1\ \algorithmicdo}

\newlist{clustered}{itemize}{2}
\setlist[clustered]{label={$\blacktriangleright$},leftmargin=*}

\date{}

\begin{document}

\conferenceinfo{DIR}{'16, November 25, 2016, Delft, The Netherlands}
\setcopyright{rightsretained}

\title{Topical Generalization for Presentation of User Profiles}

%
%
%
%
%

\numberofauthors{1} 
%
\author{
\centerline{
Alex Olieman $^{a,b}$
~ Jaap Kamps $^b$
~ Gleb Satyukov $^{a,c}$
~ Emil de Valk $^a$} \\ 
\sharedaffiliation
\affaddr{\mbox{}$^a$ Stamkracht, Amsterdam, The Netherlands}\\
\affaddr{\mbox{}$^b$ University of Amsterdam, Amsterdam, The Netherlands}\\
\affaddr{\mbox{}$^c$ Leiden University, Leiden, The Netherlands}\\
\affaddr{\normalsize
alex@stamkracht.com
~ kamps@uva.nl
~ gleb@stamkracht.com
~ emil@stamkracht.com}
}

\maketitle

\begin{abstract}
Fine-grained user profile generation approaches have made it increasingly feasible to display on a profile page in which topics a user has expertise or interest. Earlier work on topical user profiling has been directed at enhancing search and personalization functionality, but making such profiles useful for human consumption presents new challenges.
With this work, we have taken a first step toward a semantic layout mode for topical user profiles. We have developed a topical generalization approach which finds coherent groups of topics and adds labels to them, based on their association with broader topics in the Wikipedia category graph. A nested layout mode, employing topical generalization, is compared with a simpler flat layout mode in our user study. The results indicate that users favor the nested structure over flat profiles, but tend to overlook the specific topics on the lower level. We propose a third layout mode to address this issue.
\end{abstract}



\section{Introduction}
\label{sec:introduction}
A widely studied aspect of user profiling deals with the relations between users and topics. In an enterprise context, the emphasis has traditionally been on ``expertise'' relations, uncovering who is knowledgeable or skilled in which topics. Such relations, represented in expertise profiles, have powered expert finding applications for decades \cite{Guy2013}. Topical user profiles that model ``interest'' relations have been more prevalent on the public social web, where they are mainly leveraged for personalization purposes. This line has been blurred by the adoption of social media functionality into enterprise information systems \cite{Guy2013}.

Recent advances in user interest profiling have demonstrated the feasibility of automatically generating up-to-date profiles by identifying fine-grained topics (e.g. named entities) in user contributions, and linking them to canonical topic representations on the Semantic Web. Most application-oriented work in this area has focused on technology to enhance clustering, recommendation, and search functionality \cite{Kapanipathi2014,Michelson2010,Orlandi2012}. These use-cases, however, all deal with the utility of interest profiles for further computer processing, and consequently haven't raised many questions about how to display expertise or interest profiles to users.

The interaction between information seekers and user profiles is central to \textit{expertise selection}, the process of choosing an expert from a list of recommended people \cite{Yarosh2012}. Consider, for example, the need to find a suitable person to help solve a problem or to participate in a brainstorming session. After querying a system with the topic(s) related to the problem or event, the information seeker has to select one or several people to contact. The topics associated with a retrieved profile, e.g. displayed as an expertise summary or as frequently used tags, are rated by users as highly useful cues in their selection process, leading to faster selection of experts who, in turn, are more likely to respond to the seeker's inquiry and to consider themselves appropriate for what is asked of them \cite{Yarosh2012}.
%
%
%

In the broader domain of enterprise \textit{people search} there are also less task-oriented scenarios that can benefit from the availability of information about a person's expertise and interests. For instance, to find out what else the author of an interesting blog post has written about, or for newly hired employees to familiarize themselves with existing department members. Fine-grained profile generation approaches have made it increasingly feasible to present this information on a profile page as a collection of topic labels. These topical profiles, however, can grow in size rapidly as a consequence of user activity, which leads to a new challenge: \textit{how to display a large amount of topics in a user-friendly manner?}

We see two main obstacles to overcome in order to turn a topical user profile into an effective decision-making tool: (1) topics are taken out of their original context and need to be recontextualized to describe a person instead of a document (e.g. it will be easier to understand what it means that Jane writes about `anodizing,' `electrowinning,' and `salt spray tests,' if we are also told that she is a metallurgical process expert), and (2) there can be much variation in specificity between the topics within a profile (e.g. as specific as `Glastonbury Festival 2008' and as general as `society').

In this paper, we investigate the possibility of using a predefined knowledge organization system to add semantic structure to the layout of topical user profiles. First, we describe how we identify topics in user-generated content and subsequently aggregate these topics to form topical user profiles. We propose a secondary step, topical generalization, that has the goal of improving the usability of profile pages by finding and labeling semantically coherent groups of topics. Our proof-of-concept makes use of DBpedia's derivative of the Wikipedia category graph, and its utility is evaluated by means of a user study. Finally, we interpret the results of the user study, clarify a potential pitfall of the nested layout mode, and suggest how it can be avoided.

\newpage

\section{Linking users to topics}
\label{sec:linking}
A common point of departure for generating topical user profiles is to identify topics in individual user contributions, to subsequently aggregate these topics into profiles based on the interactions between users and content (e.g. authored, commented, edited, liked). Our chosen method of topics identification goes beyond keyword extraction by disambiguating the topics that are found in text into DBpedia resources. This entity recognition and disambiguation (ERD) approach helps to overcome issues with polysemy and homonymy \cite{Kapanipathi2014,Michelson2010}. As an additional benefit, it greatly facilitates the interoperability of user profiles by linking to topic representations that are accessible through public URIs \cite{Orlandi2012}.

Our approach for generating profiles from user contributions is as follows: First take the set of source documents that are associated with an individual user, find the substrings that mention topics, disambiguate them, and link them to DBpedia. We use DBpedia Spotlight \cite{Daiber2013} to perform this ERD step. Subsequently count how many source documents link to each of the topics, and store links between the user and the aggregated topics that are weighted by the (document) inlink counts of the topics. This approach is similar to ``resource-based profiling'' in \cite{Orlandi2012}, which uses an additional time decay function on topic weights.

As a consequence of using DBpedia as the target Knowledge Base in the ERD step, the identified topics can represent concepts and entities that are much more specific than the terms in which we would describe our own areas of expertise and interest. Existing approaches address this issue by viewing the fine-grained ERD outputs as intermediary, and the Wikipedia categories they are included in as the final profile topics \cite{Kapanipathi2014,Michelson2010,Orlandi2012}. Evaluation with users, however, indicates that profiles that only consider categories as topics are less precise than profiles that consist of more specific topics \cite{Orlandi2012}. We, therefore, use only the ERD output for profile generation, and propose to use the Wikipedia category graph in a separate step, to display topics on a profile page.

\section{Topical generalization}
\label{sec:topical-generalization}
The goal of our topical generalization approach is to find semantically coherent groups of topics in user profiles, and to pair these groups with categories that convey a context with which the topics are associated. This is not equivalent to a classification problem, because the Wikipedia category graph is a folksonomy in which multiple views on how the world is categorized co-exist. Whereas in a taxonomy objects of different kinds would only be distantly related, e.g.: `emerald (mineral),' `necklace (artifact),' and `pearl (organic object),' in Wikipedia they may be more closely related due to a shared context (e.g. the `Jewellery' category).

\subsection{Pairing Groups of Topics with Categories}

Let $A$ be the set of DBpedia URIs that are derived from Wikipedia articles, and $B$ be the set of URIs derived from Wikipedia categories ($A \cap B = \emptyset$). Let $E \subset A$ be the set of topics in a user profile. For each $e \in E$, find its parent categories and their broader categories $C_{e} \subset B$, traversing up to $m$ edges in the graph. In DBpedia the edges from articles to categories are represented as \code{dct:subject}\footnote{For namespace prefixes, see \url{https://dbpedia.org/sparql?nsdecl}.}, and from categories to their superordinates as \code{skos:broader}. Together, these categories form the set
\begin{equation}
C = \bigcup_{e \in E} C_{e}
\end{equation}

The traversal which finds $C_{e}$ can be implemented in several ways, depending on the way in which the category graph is accessed (e.g. by loading DBpedia dumps into a triplestore or graph database). Query 1 provides an executable example of this traversal as a SPARQL\footnote{SPARQL -- \url{https://www.w3.org/TR/sparql11-query/}} query. Throughout this paper we use $m=3$.

\begin{lstlisting}[
	basicstyle=\fontsize{7.5}{8}\tt, 
    language=SPARQL, 
    float, 
    floatplacement=H, 
    caption={Example SPARQL query to find $C_e$ with $m=3$. This example can be executed on \url{http://dbpedia.org/snorql}.}, 
    label=traversal,
    %frame=leftline,
    %xleftmargin=3.4pt,
    %rulecolor=\color{gray},
    abovecaptionskip=-4pt,
    belowskip=-0.5em]
SELECT DISTINCT ?c0 ?c1 ?c2 WHERE {
  VALUES ?e { dbr:Pearl } .
  {
    ?e dct:subject ?c0.
  } UNION {
    ?e dct:subject/skos:broader ?c1.
    FILTER NOT EXISTS {
      ?e dct:subject ?c1.
    }
  } UNION {
    ?e dct:subject/skos:broader
      /skos:broader ?c2.
    FILTER NOT EXISTS {
      ?e dct:subject ?c2.
    } .
    FILTER NOT EXISTS {
      ?e dct:subject/skos:broader ?c2.
    }
  }
}
\end{lstlisting}

In order to find representative categories for groups of topics, let $D = (d_{ce}: c\in C, e\in E)$ be a $|C|\times |E|$ sparse matrix which contains category--topic distances $d_{ce}$, where $d_{ce}$ is the length of the \code{skos:broader} path between $e$ and $c$, and $0 \leq d_{ce} < m$. To continue the SPARQL example, in which $e =$ \code{dbr:Pearl}, we find (among many others): $d_{dbc:Gemstones,e} = 0$, and $d_{dbc:Materials,e} = 2$. $D$ is initialized with $null$ values, and has the index set $C$ for rows and $E$ for columns.

Our aim is to identify categories that have many transitive inlinks from topics, with the shortest possible path distances to avoid overgeneralization. This intuition is similar to the ``Intersect Booster'' from \cite{Kapanipathi2014}. To order $C$ by suitability to represent a group of topics, we define:
\begin{equation}
AdoptionRank(c, D) = \gamma + \frac{\sum_{e\in E} d_{ce}}{Coverage(c, D)}
\end{equation}
where
\begin{align*}
Coverage(c, D) & = |(d_{ce}: d_{ce} \neq null, e\in E)|  \\
\gamma & = \frac{\kappa}{Coverage(c, D)^2}
\end{align*}

A lower $AdoptionRank$ value, following its definition, indicates that a category is superordinate to a larger group of topics and/or stands in a more direct relation to these topics. The penalty $\gamma$ is applied to reduce the likelihood of ties, and $\kappa$ is a constant for which higher values favor the number of inlinks over the row-wise sum of $d_c$. In this paper we use $\kappa=1$.

\begin{algorithm}
\captionof{algorithm}{Cluster topics under categories.} \label{alg:cluster-entities}
\begin{algorithmic}[1]
\Require $\Phi$, an iterator over sorted rows of $D$
\State $toAssign \gets E$
\State $clusters \gets \emptyset$

\ForEach {$c,d_c \in \Phi$}
	\State $inBoth \gets toAssign \cap \textup{indices}(d_c)$
    \If {$ \left\vert{inBoth}\right\vert > 1 $}
    	\State $clusters \gets clusters \cup \{ \langle c, \textup{indices}(d_c) \rangle \}$
        \State $toAssign \gets toAssign \setminus inBoth$
        \If {$toAssign = \emptyset$}
        	\State \textbf{break}
        \EndIf
    \EndIf
\EndFor
\State $orphans \gets toAssign$
\State \Return $clusters, orphans$
\end{algorithmic}
\end{algorithm}

\subsection{Cluster Selection}
\label{sec:layout}
The ranked sequence of category--topic-group pairs is not meant to be displayed directly to users. Each topic is included in multiple groups, and we reduce this redundancy by transforming the pairs into labeled clusters. Algorithm \ref{alg:cluster-entities} iterates over the sorted pairs and ``promotes'' the pair to a cluster if it includes at least two topics that have not been assigned to previous clusters.

It is likely, if the set of input topics is sufficiently large, that some groups are completely subsumed by other groups of topics. This could easily be exploited by the clustering algorithm to create a hierarchy of clusters whenever possible. We choose to leave this possibility for future work. 

\newpage

\section{User study}
\label{sec:user-study}

A between-group experimental design is used to evaluate whether the layout mode of profile topics has an effect on users' accuracy, error diversity, and task duration, when they are instructed to assign profiles to top-level categories.
In this study we compare a flat list layout with a nested list layout that makes use of our topical generalization approach. The flat layout is a straightforward presentation of profile topics which requires little effort in terms of user interaction, but does not show coherence between the topics.

The design of the nested layout is partly based on suggestions that were found in related work. It features a combination of entities and categories, and similar topics are clustered \cite{Orlandi2012}. It is also compact, by collapsing the profiles into a few broader topics \cite{Michelson2010}. Consequently, it offers the user more information in the form of structure and added category labels, but also requires more effort to expand the clusters in order to view their underlying topics.

\subsection{Participants}
\label{sec:participants}
Sixty-four participants, aged 20-61 ($\mu$=31.8; SD=9.7), of which 26 female, completed our experimental procedure. A large majority of participants was highly educated. They were recruited via convenience sampling by an invitation that was spread through online social networks, originating from the authors and at least eight of their co-workers, and was shared from there on. Participation was limited to speakers of the Dutch language. This language was preferred over English because of its high percentage of native speakers, so as to reduce the risk of language comprehension as a confounder.

\subsection{Task}
\label{sec:task}

In order to measure the variables of interest, we implemented a manual classification task for our remote user study. This task was framed as an expert finding scenario in which the participants were looking for journalists with various specializations. The hypothetical search system, however, did not allow for a more specific query than ``journalist.'' This was used as an excuse to let the participants evaluate a sequence of user profiles, which they needed to inspect in order to classify them into one of five top-level categories: Art, Finance, Sports, Technology, Travel, or into an ``Other'' category.

\begin{figure}[ht]
	\includegraphics[width=\columnwidth]{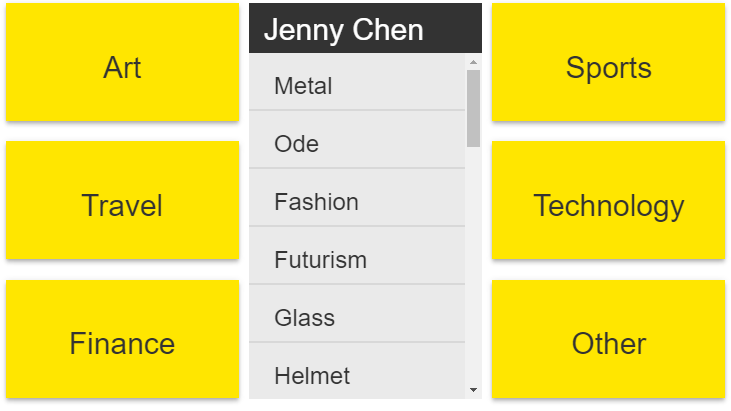}
	\caption{An ``Other'' profile in the control condition.}\label{fig:ddct-flat}
\end{figure}
\begin{figure}[ht]
	\includegraphics[width=\columnwidth]{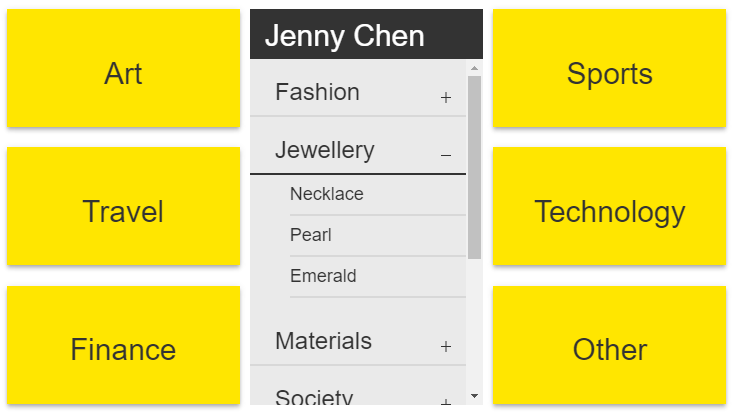}
	\caption{An ``Other'' profile in the experimental condition.}\label{fig:ddct-nested}
\end{figure}

In the control condition participants were presented with regular user profiles, as described in Section \ref{sec:linking}, displayed as a flat list of the specific topics that were extracted from the source documents (see Figure \ref{fig:ddct-flat}). The topics were ordered from highest to lowest indegree, and ties were resolved alphabetically on the topic labels.

In the experimental condition participants were asked to classify profiles with an added upper level of broader categories (i.e. generalized topics). The clusters that are produced by the approach in Section \ref{sec:topical-generalization} are sorted by the sum of their source document counts (i.e. indegrees), from high to low. The topics that aren't members of any cluster (i.e. orphans) are added at the bottom. The clusters are laid out on the profile page as a nested list, with category labels on the top level, and the underlying topics on the second level, as shown in Figure \ref{fig:ddct-nested}. This type of widget is commonly known as an ``accordion,'' which is initially in a collapsed state.

\subsection{Profiles}
\label{sec:fictional-profiles}

We have created 18 profiles of fictional journalists--three per target category--that were arranged in a sequence that was randomized prior to opening the study to participants. The source documents for the profiles were selected by searching within the sites of Dutch news publishers that are geared towards the given categories. These sites were searched with terms that indicate demarcated sub-topics within each category. For each profile, 3-5 source documents were selected (depending on length), preferably by the same author. This resulted in profiles consisting of 5-94 topics ($\mu$=30.2). We added fictional names to the profiles to reinforce the idea that these were profiles of different users, without giving away clues about the correct category.

\subsection{Experimental Procedure}
\label{sec:experimental-procedure}
Participants were shown one profile at a time, which they dragged and dropped into the target category that best matched their interpretation of the profile. The categories were displayed as relatively large rectangles positioned on both sides of the profile. During this task we recorded the duration between when the profile was displayed and when it was classified, as well as into which category the profile was dropped.
After the task was completed, participants landed on a page where they were asked to indicate their (dis)agreement with three statements (1=\textit{completely disagree} to 5=\textit{completely agree}): \textit{the names of the topics were clear}, \textit{the topics were arranged usefully}, and \textit{the topics gave a clear picture of what kind of journalist the profile belonged to}. Finally, participants were invited to add any remarks about the study.

\subsection{Results}
\label{sec:results}

Both conditions exhibited a significant learning effect in the time spent with each subsequent profile. To account for this in our analysis, we divide our observations into a warm-up phase (profiles 1-4; control $R^2 = 0.42$, experimental $R^2 = 0.27$)\footnote{Coefficient of determination $R^2$ for $\ln(\Delta t) = \alpha + \beta \times task\_id$} and a testing phase (profiles 5-18; control $R^2 = 0.08$, experimental $R^2 = 0.02$). In the warm-up phase participants in the control group classified the profiles significantly more accurately, whereas the experimental group spent significantly more time, which indicates that it was easier for participants to learn to use the flat layout.

Once the participants were familiarized, however, the observed effect of layout mode on task performance was subtler. Participants who were presented with flat profiles on average classified them slightly more accurately (0.76$\pm$0.05) than participants who saw the nested profiles (0.73$\pm$0.05). The difference in average time spent with each profile was minor, being 7.51$\pm$0.47 s in the control group, versus 7.97$\pm$0.65 s in the experimental group. This suggests that participants in the experimental group did not feel the need to spend much time interacting with the generalized topics that were available in their nested layout.

Use of the nested layout did lead to significantly more diverse classification errors than the flat layout. We took, for each profile and per condition, the proportional abundance of the incorrect guesses, and computed the Shannon entropy $H'$ of this distribution as a diversity index. A pairwise comparison of the resulting diversity values is found to be significant with $p<0.05$ by a Wilcoxon signed-rank test.

The observations furthermore exhibit a task effect between the profiles regarding accuracy and duration. Six of the 14 profiles were classified more accurately by the control group, versus three profiles by the experimental group. Both groups each classified one profile significantly faster than the other group did.

The reflection of participants on the given task indicates that the experimental group found the profiles less usable overall. The topic names were judged to be clear by 61\% of the control group, versus 48\% of the experimental group. For 35\% of the control group the profiles gave a clear picture of the represented persons, whereas this was only the case for 18\% of the experimental group. The flat presentation of profiles, however, was rated as not useful by 61\% of participants, while 48\% of participants who worked with the nested profiles disagreed that the topics were arranged usefully.

\section{Discussion and Outlook}
\label{sec:discussion}
With this work, we have taken a first step toward a semantic layout mode for topical user profiles. We have developed a topical generalization approach which forms clusters of topics based on their association with broader topics in the Wikipedia category graph. A nested layout mode, employing topical generalization, is compared with a simpler flat layout mode in our user study.

The results of our experiment suggest that the nested layout does not sufficiently compel users to inspect the lower level of generalized topics.  Some users may not have found the additional information worth the effort of interaction, but others were unaware that they could do so. Several participants reported that they didn't read the entire introductory text, and two participants remarked that they only found out that they could expand the generalized topics near the end of the task. This is unfortunate because category labels are inherently more abstract than their underlying topics. A profile that is perceived to consist only of general categories has limited utility, because superordinate categories are less cognitively accessible and less informative than ``basic-level'' categories \cite{Tanaka1991}.  Being unaware of the specific topics an expert engages with, makes users prone to overgeneralizing the expert's area of expertise.

A per-profile analysis of differences between the groups shows, however, that some profiles were easier to classify using the nested layout. Moreover, the structure of the nested profiles was found to be useful by more participants than that of the flat profiles.

\begin{figure}[ht]
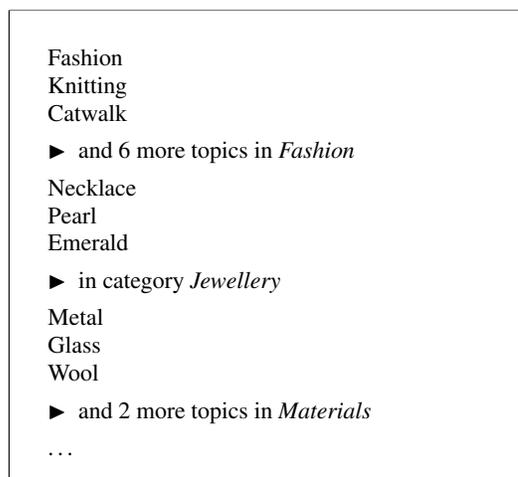

\centering
\vspace*{-10pt}
\caption{Example of proposed clustered profile layout.}
\label{fig:clustered_profile}
\end{figure}

We would like to investigate, in future research, whether another layout mode can benefit from both specificity and compactness. We propose to switch the foreground and background of the generalized topics, by displaying the $k$  topics with the highest indegrees within each cluster, and using their category label to provide context without obfuscating them. A mock-up of such a clustered layout is shown in Figure \ref{fig:clustered_profile}. The indented lines below the top-$k$ topics would, by our design, reveal all topics in the cluster when they are clicked.

The proposed layout mode has recently been incorporated into a social media platform that is used within several organizations, with the aim of evaluating it longitudinally with real user behavior.

\begin{mdframed}[
	innermargin=20,
    outermargin=22,
    innertopmargin=10,
    innerbottommargin=10,
    skipabove=10
]
\begin{itemize}[
	noitemsep,
    label={},
    leftmargin=*,
    topsep=0pt
]

\item Fashion
\item Knitting
\item Catwalk
\begin{clustered}
  \item and 6 more topics in \emph{Fashion}
\end{clustered}

\item Necklace
\item Pearl
\item Emerald
\begin{clustered}
  \item in category \emph{Jewellery}
\end{clustered}

\item Metal
\item Glass
\item Wool
\begin{clustered}
  \item and 2 more topics in \emph{Materials}
\end{clustered}

\item \ldots

\end{itemize}
\end{mdframed}

\vspace*{-10pt}
\bibliographystyle{abbrv}
\bibliography{main}

\end{document}